\documentclass{aip-cp}

\usepackage[numbers]{natbib}
\usepackage{rotating}
\usepackage{graphicx}
\usepackage{wrapfig}

\begin{document}
\title{Positron trapping and annihilation at interfaces between matrix and 
cylindrical or spherical precipitates modeled by diffusion-reaction theory }

\author[aff1]{Roland W\"{u}rschum\corref{cor1}}
\author[aff1]{Laura Resch}
\author[aff1]{Gregor Klinser}
\affil[aff1]{Institute of Materials Physics, Graz University of Technology, Petersgasse 16, A-8010 Graz, Austria}
\corresp[cor1]{Corresponding author: wuerschum@tugraz.at}

\maketitle

\centerline{\em \large Dedicated to Professor Alfred Seeger} 

\vspace{.3cm}

\begin{abstract}
The exact solution of a diffusion$-$reaction model for the
trapping and annihilation of positrons at interfaces of precipitate$-$matrix composites 
is presented considering both cylindrical or spherical precipitates.
Diffusion-limitation is taken into account for interfacial trapping  from the surrounding matrix
as well as from the interior of the precipitate.
Closed-form expressions are obtained for the mean positron lifetime and
for the intensity of the positron lifetime component associated with
the interface-trapped state.
The model contains as special case also positron trapping at
extended open-volume defects like spherical voids or hollow cylinders.
This makes the model applicable to all types
of cylindrical- and spherical-shaped extended defects irrespective of their size and  their number density.
\end{abstract}

\section{Introduction}
Positron ($e^+$) annihilation techniques are nowadays widely applied to study structurally complex materials. Here, a composite structure consisting of a crystalline matrix with embedded precipitates is of particular application relevance. 
It is, however, well known that $e^+$ trapping at such extended defects like precipitates cannot be correctly described
by standard rate theory but demands for analysis in the framework of diffusion-reaction theory. 
Whereas diffusion-limited $e^+$ trapping at grain boundaries of crystallites (or equivalently at surfaces of particles)
has been quantatively modeled by several groups 
\cite{Dupasquier93, Wuerschum96, dryzek1998, dryzek1999, Oberdorfer09},
diffusion-limited $e^+$ trapping at interfaces of precipitate$-$matrix composites is more complex and has not been treated until recently \cite{Wuerschum18}. In addition to diffusion-limited interface trapping 
from the interior of the precipitate, in particular the interfacial trapping from the surrounding matrix
has to be treated taking into account diffusion limitation.

Following our earlier publications on grain boundaries 
\cite{Wuerschum96,Oberdorfer09}, in the present work the diffusion-reaction limited $e^+$ trapping 
at interfaces of precipitate$-$matrix composites is mathematically handled by means of Laplace transformation,
which yields closed-form expressions for the mean $e^+$ lifetime and for the intensity of the annihilation component associated with the interfacial trapped state. These solutions can be conveniently applied for the analysis of experimental data. The present work treats the composite structure both for cylindrical- and for spherical-shaped precipitates. The solutions of the spherical-symmetric model were already reported recently 
in a broader context including modeling of voids and small clusters \cite{Wuerschum18}.  

%%%%%%%%%%%%%%%%%%%%%%%%%%%%%%%%%%%%%%%%%%%%%%%%%%%%%%%%%%%%%%%%%%%%%%%%%%%%%%%%%%%%%%%%%%%%%%
\section{Cylindrical and spherical diffusion$-$reaction  model}
\begin{wrapfigure}[16]{r}[0cm]{5cm}
\mbox{\includegraphics[width=5cm,angle=0]{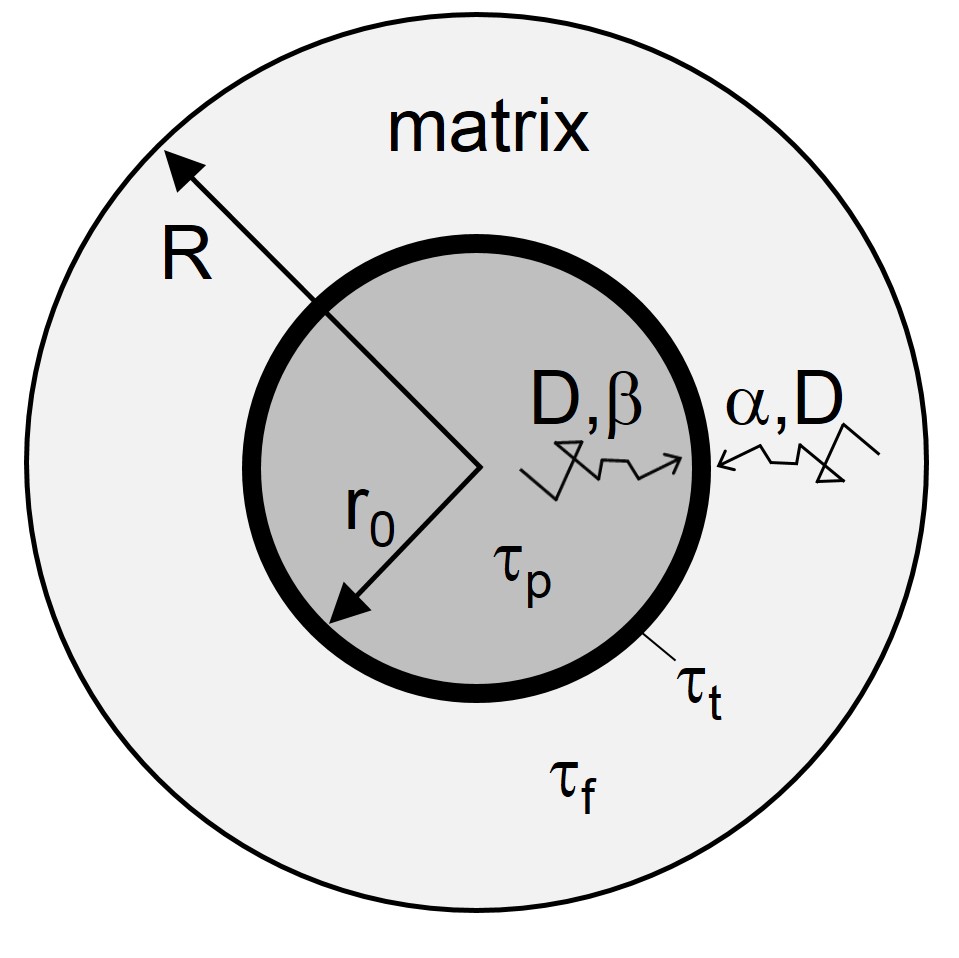}}
\caption{Geometry of diffusion-reaction model.} 
\label{fig:1}
\end{wrapfigure}

The geometry of the diffusion-reaction model is schematically shown in Fig.~\ref{fig:1}.
The model describes positron ($e^+$) trapping and annihilation in a precipitate$-$matrix composite with either cylindrical- or spherical-shaped precipitates of radius $r_0$. Positron annihilation from the free bulk state is considered both for the
matrix and for the precipitate, each characterized by a specific free $e^+$ lifetime, denoted $\tau_f$ and $\tau_p$, respectively. 
Positron trapping into the precipitate$-$matrix interface is considered to be diffusion- and reaction-limited 
for both the trapping from the surrounding matrix and from the interior of the precipitate. 
Trapping from the matrix is characterized by the specific trapping rate $\alpha$ and the $e^+$ diffusion coefficient $D$ 
and from the precipitate by the rate $\beta$ and the same value of $D$. The number density $N_p^{sphere}$ of spherical precipitates per unit volume or that  of cylindrical precipitates ($N_p^{cyl}$) per unit area 
is related to the outer radius $R$ of the surrounding matrix: 
\begin{equation}
\label{eq:N_p}
N_p^{sphere} = \Bigl( \frac{3}{4 \pi R^3} \Bigr)^{1/3} \, , \hspace{1cm} N_p^{cyl} = \Bigl( \frac{1}{\pi R^2} \Bigr)^{1/2} \, .
\end{equation}

Since detrapping of $e^+$ from the interface is neglected, the two $e^+$ trapping processes into the precipitate$-$matrix interface from inside the precipitate and from the surrounding matrix are completely decoupled apart from the initial condition.
Both processes can, therefore, be treated independently. This implies further  that diffusion-reaction limited $e^+$ trapping from inside the precipitates into the interfaces can be treated completely analogous to the corresponding models of $e^+$ trapping at grain boundaries of spherical \cite{Wuerschum96, Oberdorfer09} or cylindrical crystallites \cite{dryzek1998}.

The temporal and spatial evolution of the density  $\rho_m$ of free $e^+$  in the matrix 
and of those  in the precipitates ($\rho_p$) is governed each by the diffusion equation: 
\\
\begin{equation}
\label{eq:diffusion}
\frac{\partial \rho_{m/p}}{\partial t}=
D\nabla^{2}\rho_{m/p}-  \frac{\rho_{m/p}}{\tau_{f/p}}
\end{equation}
where $\tau_f$ and $\tau_p$ denote  the above mentioned free $e^+$ lifetimes.
The $e^+$ density $\rho_t$ in the trapped state of the precipitate$-$matrix interface consists of two parts, i.e.,
the densities $\rho^{(m)}_t$ and $\rho^{(p)}_t$ due to trapping from the matrix and the precipitate, respectively.
These densities obey the rate equations
\begin{equation}
\label{eq:rho_t}
\frac{\mathrm{d}\rho^{(m/p)}_t}{\mathrm{d}t}=
\alpha/\beta \, \rho_{m/p}(r_{0},t)-\frac{1}{\tau_t}\rho_t \, ,
\end{equation}
with the specific trapping rates  $\alpha$, $\beta$ (unit ms$^{-1}$) defined above.

The continuity of the ${\rm e}^{+}$ flux at the boundary 
between the matrix ($m$) and the interface or between the precipitate ($p$) and the interface
is expressed by
\begin{eqnarray}
\label{eq:inner_bound}
D\nabla \rho_m\Big|_{r=r_{0}}-\alpha \rho_m(r_{0},t)=0 \, , \quad 
D\nabla \rho_p\Big|_{r=r_{0}}+\beta \rho_p(r_{0},t)=0 \, .
\end{eqnarray}

The outer boundary condition
\begin{equation}
\label{eq:outer_bound}
\frac{\partial \rho_m}{\partial r}\Big|_{r=R}=0
\end{equation}
reflects the vanishing $e^+$ flux through the outer border ($r = R$) of the diffusion sphere. 
As initial condition a homogeneous distribution of $e^+$  in the matrix and the 
precipitate ($\rho_m (0) = \rho_p (0)$) is assumed without  $e^+$ in the trapped state ($\rho_t (0) = 0$) for $t=0$.

In order to obtain closed-form solutions, 
the time dependence of the diffusion and rate equations (\ref{eq:diffusion}, \ref{eq:rho_t})
is handled by means of Laplace transformation. 
As briefly outlined in the appendix,  solving the $r$-dependent differential equation and
subsequent volume integration,  finally leads to the Laplace transform $\tilde{n}(p)$ of  
the total probability $n(t)$ that a $e^+$ implanted at $t=0$ has not yet been annihilated at time $t$.
$\tilde{n}(p)$ contains the entire information of the $e^+$ annihilation characteristics (see next section).
For the cylindrical case, $\tilde{n}(p)$ reads:
\begin{eqnarray}
\tilde{n}(p)=\Bigl(\frac{r_0}{R}\Bigr)^2 \frac{1}{\tau_p^{-1}+p}
\Biggl\{1 + \frac{2 \beta}{r_0} \, \frac{\tau_p^{-1} - \tau_t^{-1}}{\tau_t^{-1}+p} \, \frac{\Theta}{\gamma' 
(\Theta \gamma' D + \beta) } \Biggr\}+ 
\frac{1}{\tau_f^{-1}+p}
\Biggl\{1 - \Bigl(\frac{r_0}{R}\Bigr)^2 + \frac{2 \alpha r_0}{R^2} \, \frac{\tau_f^{-1} - \tau_t^{-1}}{\tau_t^{-1}+p} \,
\frac{\Lambda_1}{\gamma ( \Lambda_1  \gamma D - \alpha \Lambda_0)} \Biggr\}
\label{Laplace_n_cyl}
\end{eqnarray}
with 
\begin{equation}
\Theta=\Theta(\gamma' r_0)=\frac{I_1 (\gamma' r_0)}{I_0 (\gamma' r_0)}  \, , \quad
\quad \Lambda_{0/1}= \Lambda_{0/1} (\gamma r_0, \gamma R)= I_{0/1}(\gamma r_0)  K_1(\gamma R) +/- 
 K_{0/1}(\gamma r_0) I_1(\gamma R) \, ,
\label{eq:L_12}
\end{equation}
and
\begin{equation}
\label{eq:gamma'}
\gamma^{2} = \frac{\tau_{f}^{-1}+p}{D} \, , \hspace{0.3cm}
\gamma'^{2} = \frac{\tau_{p}^{-1}+p}{D} \, .
\end{equation}
$I_{j}$, $K_{j}$ ($j = 0, 1$) denote the modified Bessel functions  \cite{olver2010nist}.
The solution for the spherical case reads 
\begin{eqnarray}
\nonumber
\tilde{n}(p) = 
\Bigl(\frac{r_0}{R}\Bigr)^3 \frac{1}{\tau_p^{-1}+p}
\Biggl\{1 + \frac{3 \beta}{r_0} \, \frac{\tau_p^{-1} - \tau_t^{-1}}{(\tau_t^{-1}+p)(\tau_p^{-1}+p)} \,
\frac{\gamma' D L(\gamma' r_0)}{\gamma' D L(\gamma' r_0) + \beta} \Biggr\}+ 
 \\
 \frac{1}{\tau_f^{-1}+p} \Biggl\{ 1-\Bigl(\frac{r_0}{R}\Bigr)^3+
\frac{3\alpha r_0^2}{R^3} \,
\frac{\tau_f^{-1}-\tau_t^{-1}}{(\tau_t^{-1}+p)(\tau_f^{-1}+p)} \,
\frac{\gamma \hat{R}-\tanh(\gamma \hat{R}) [1-\gamma^2 r_0 R]}{\gamma \hat{R}-\tanh(\gamma \hat{R}) [1-\gamma^2 r_0 R]+ \frac{\alpha r_0}{D}[\gamma R- \tanh(\gamma \hat{R})]}
 \Biggr \} 
\label{Laplace_n_sphere}
\end{eqnarray}
with $\hat{R} = R - r_0$
and the Langevin function
$L (z) = \coth z - 1/z$.

%%%%%%%%%%%%%%%%%%%%%%%%%%%%%%%%%%%%%%%%%%%%%%%%%%%%%%%%%%%%%%%%%%%%%%%%%%%%%%%
\section{\label{Results}Results for analysis of measurements}
The \underline{mean positron lifetime} $\overline{\tau}$ is obtained  by
taking the Laplace  transform  [Eq. (\ref{Laplace_n_cyl}) and (\ref{Laplace_n_sphere})] at $p = 0$
($\overline{\tau} = \tilde{n}(p=0)$).
For the precipitation$-$matrix composite with cylindrical symmetry the mean positron lifetime reads 
\begin{eqnarray}
\label{eq:tauq_cyl}
\overline{\tau} = \Bigl(\frac{r_0}{R}\Bigr)^2 \tau_p
\Biggl\{1 +  \frac{2 \beta}{r_0} (\tau_t-\tau_p) \, \frac{\Theta}{(\tau_p/D)^{1/2} \beta +  \Theta]} \Biggr\}+  % \nonumber \\ 
\tau_f
\Biggl\{1 - \Bigl(\frac{r_0}{R}\Bigr)^2 + \frac{2 \alpha r_0}{R^2} (\tau_t - \tau_f)
\frac{\Lambda_1}{\Lambda_1 - (\tau_f/D)^{1/2} \alpha \Lambda_0} \Biggr\}
\end{eqnarray}
with
$\Theta(\gamma'_0 r_0)$ and $\Lambda_{0,1}(\gamma_0 r_0, \gamma_0 R)$ according to eq.~(\ref{eq:L_12}) with $\gamma'_0 = (\tau_p D)^{-1/2}$ 
and $\gamma_0 = (\tau_f D)^{-1/2}$.
Likewise the solution of the spherical precipitates reads \cite{Wuerschum18}
\begin{eqnarray}
\label{eq:tauq_sphere}
\overline{\tau}= 
\displaystyle{\Bigl(\frac{r_0}{R} \Bigr)^3 \tau_p \Biggl\{ 
1 +  \frac{2 \beta}{r_0}(\tau_t-\tau_p) \,
\frac{L(\gamma'_0 r_0)}{(\tau_p/D)^{1/2} \beta + L (\gamma'_0 r_0) }\Biggr\}} + 
\nonumber \\
\tau_f \Biggl\{\displaystyle{1 - \Bigl(\frac{r_0}{R} \Bigr)^3 } + 
 \displaystyle{\frac{3 \alpha r_0^2}{R^3} 
\, (\tau_t-\tau_f) \, 
\frac{\gamma_0 \hat{R}-\tanh(\gamma_0 \hat{R}) [1-\gamma_0^2 r_0 R]}{\gamma_0 \hat{R}-\tanh(\gamma_0 \hat{R}) [1-\gamma_0^2 r_0 R]+ \frac{\alpha r_0}{D}[\gamma_0 R- \tanh(\gamma_0 \hat{R})]} \Biggr \} } \, .
\end{eqnarray}

The \underline{positron lifetime spectrum}
follows from  $\tilde{n}(p)$ [Eq. (\ref{Laplace_n_cyl}) and (\ref{Laplace_n_sphere})] 
by means of Laplace inversion. The single poles $p = - \lambda_i$
of  $\tilde{n}(p)$ in the complex $p$ plane define the decay rates with the relative intensities $I_i$ 
of the $e^+$ lifetime spectrum.
As usual for this kind of problem, the annihilation from the free state in the matrix and the precipitates is characterized by 
series of fast decay rates $p = - \lambda_i (i=0,1,2,\dots)$ ($\lambda_i > \tau_{f/p}^{-1}$) which follow from 
the first-order roots of $n(p)$. The rates are given by the solutions of the transcendental equations
which read for the precipitates with cylindrical or spherical symmetry  
\begin{equation}
\gamma'^{\star} \frac{J_1 (\gamma'^{\star} r_0)}{J_0 (\gamma'^{\star} r_0) }= \frac{\beta}{D} \, , \mbox{ or } \quad
\gamma'^{\star} r_0 \coth(\gamma'^{\star} r_0) = 1 - \frac{\beta r_0}{D} 
\label{eq:trans_precip}
\end{equation}
respectively, with 
$\gamma'^{\star 2} = (\lambda_{i}-\tau_p^{-1})/D$ 
and for the matrix in the cylinder- or sphere-symmetrical case
\begin{eqnarray}
\gamma^{\star} \Bigl\{ Y_1 (\gamma^{\star} r_0) J_1 (\gamma^{\star} R) - J_1 (\gamma^{\star} r_0) Y_1 (\gamma^{\star} R) \Bigr\}= 
- \frac{\alpha}{D} 
\Bigl\{ Y_0 (\gamma^{\star} r_0) J_1 (\gamma^{\star} R) - J_0 (\gamma^{\star} r_0) Y_1 (\gamma^{\star} R) \Bigr\}
 , 
\nonumber
\\
\mbox{ or } \quad \tanh(\gamma^{\star} \hat{R}) 
= \frac{\gamma^{\star} (\alpha r_0 R + D \hat{R})}{D(1+\gamma^{\star 2} r_0 R) + \alpha r_0} \hspace*{3cm}
\label{eq:trans_matrix}
\end{eqnarray}
respectively, with 
$\gamma^{\star 2} = (\lambda_{i}-\tau_f^{-1})/D$ and the Bessel functions
$Y_{j}$, $J_{j}$ ($j = 0, 1$)   \cite{olver2010nist}.

For the pole $p = - \tau_t^{-1}$ 
characterizing the interface-trapped state, $\tilde{n}(p)$ directly yields the corresponding intensity $I_t$ of this 
positron lifetime component. This intensity 
\begin{equation}
I_t = I_t^{precip} + I_t^{matrix} 
\label{eq:I_t_tot}
\end{equation}
is composed of the two parts which arise from $e^+$ trapping into the interface 
from the precipitate ($I_t^{precip}$) and from the matrix ($I_t^{matrix}$).
For the cylindrical symmetry, $\tilde{n}(p)$ (eq.~\ref{Laplace_n_cyl}) yields for the pole $p = - \tau_t^{-1}$:
\begin{equation}
I_t^{matrix} = \frac{2 \alpha r_0}{R^2} \, \frac{\Lambda_1}{(\tau_f^{-1} - \tau_t^{-1}) \Lambda_1 - \gamma_t \alpha \Lambda_0} \, ,
\label{eq:I_t_m_cyl}
\end{equation}
\begin{equation}
I_t^{precip} = \Bigl(\frac{r_0}{R}\Bigr)^2 \,
 \frac{2 \beta}{r_0} \, \frac{\Theta}{(\tau_p^{-1} - \tau_t^{-1}) \Theta + \beta \gamma'_t} 
\label{eq:I_t_p_cyl}
\end{equation}
with $\Theta=\Theta(\gamma'_t r_0)$  and 
$\Lambda_{0,1}(\gamma_t r_0, \gamma_t R)$ according to eq.~(\ref{eq:L_12}) with 
\begin{equation}
\label{eq:gamma_t}
\gamma_t^{2} = \frac{\tau_{f}^{-1} - \tau_t^{-1}}{D} \, , \hspace{0.3cm}
\gamma'^{2}_t = \frac{\tau_{p}^{-1} - \tau_t^{-1}}{D} \, .
\end{equation}
The respective total intensities arising from free $e^+$ annihilation in the precipitate and in the matrix are given by
the volume-weighted complementary values:
\begin{equation}
I_{bulk}^{precip} = \Bigl(\frac{r_0}{R}\Bigr)^2 \, (1 - I_t^{precip}) \, , \hspace{.3cm}
I_{bulk}^{matrix} = \Bigl[1- \Bigl( \frac{r_0}{R}\Bigr)^2 \Bigr] \, (1 - I_t^{matrix}) \, .
\label{eq:I_cyl_bulk}
\end{equation}

The intensities for the spherical case read \cite{Wuerschum18}
\begin{equation}
\label{eq:I_t_m_sphere}
I_t^{matrix}= \frac{3 \alpha r_0^2}{R^3} \, \frac{1}{\tau_f^{-1}-\tau_t^{-1}} \,
\frac{\gamma_t \hat{R}-\tanh(\gamma_t \hat{R}) [1-\gamma_t^2 r_0 R]}{\gamma_t \hat{R}-\tanh(\gamma_t \hat{R}) [1-\gamma_t^2 r_0 R]+ \frac{\alpha r_0}{D}[\gamma_t R- \tanh(\gamma_t \hat{R})]} \,,
\end{equation}
\begin{equation}
\label{eq:I_t_p_sphere}
I_t^{precip} = \Bigl(\frac{r_0}{R} \Bigr)^3 \, \frac{3 \beta}{r_0} \, 
\frac{1}{\tau_p^{-1}-\tau_t^{-1}} \, 
\frac{\gamma'_t D L(\gamma'_t r_0)}{
 \beta + \gamma'_t D L(\gamma'_t r_0)  } \, ,
 \,
\end{equation}
\begin{equation}
I_{bulk}^{precip} = \Bigl(\frac{r_0}{R}\Bigr)^3 \, (1 - I_t^{precip}) \, , \hspace{.3cm}
I_{bulk}^{matrix} = \Bigl[1- \Bigl( \frac{r_0}{R}\Bigr)^3 \Bigr] \, (1 - I_t^{matrix}) \, .
\label{eq:I_sphere_bulk}
\end{equation}

As long as the precipitate diameter is remarkably lower than the $e^+$ diffusion length [$(D \tau_p)^{-1/2}$] in the precipitate, 
diffusion limitation of $e^+$ trapping from the precipitate into the interface with the matrix can be neglected, so that this part of 
the $e^+$ trapping process can be reasonably well  described by standard reaction theory. In this case, the first summand 
of the mean $e^+$ lifetime in eq.~(\ref{eq:tauq_cyl}) and eq.~(\ref{eq:tauq_sphere}) simplifies to
\begin{equation}
\Bigl(\frac{r_0}{R} \Bigr)^2 \tau_t \frac{\tau_t^{-1} + \frac{2 \beta}{r_0}}{\tau_p^{-1} + \frac{2 \beta}{r_0}} \, , \mbox{ or } \quad 
\Bigl(\frac{r_0}{R} \Bigr)^3 \tau_t \frac{\tau_t^{-1} + \frac{3 \beta}{r_0}}{\tau_p^{-1} + \frac{3 \beta}{r_0}} \, ,
\label{eq:tauq_rate}
\end{equation}
respectively. In the same manner, the intensity $I_t^{precip}$ for the cylindrical  [Eq.~\ref{eq:I_t_p_cyl}] and spherical case 
[Eq.~\ref{eq:I_t_p_sphere}] simply reads, 
\begin{equation}
I_t^{precip} = \Bigl(\frac{r_0}{R} \Bigr)^2 \, \frac{2 \beta}{r_0} \,  \frac{1 }{\tau_p^{-1} + \frac{2 \beta}{r_0}- \tau_t^{-1}}  
\, , \mbox{ or } \quad
I_t^{precip} = \Bigl(\frac{r_0}{R} \Bigr)^3 \, \frac{3 \beta}{r_0} \,  \frac{1 }{\tau_p^{-1} + \frac{3 \beta}{r_0} - \tau_t^{-1}} \, .
\label{eq:I_p_rate}
\end{equation}

%%%%%%%%%%%%%%%%%%%%%%%%%%%%%%%%%%%%%%%%%%%%%%%%%%%%%%%%%%%%%%%%%%%%%%%%%%%%%%%%%%%%%%%
\section{Discussion}
The presented model with the exact solution of diffusion-reaction
controlled trapping at interfaces of matrix$-$precipitate composites with cylindrical or spherical symmetry
yields closed-form expressions
for the mean positron lifetime $\overline{\tau}$ [Eq. (\ref{eq:tauq_cyl}), (\ref{eq:tauq_sphere})]
and for the relative intensity $I_t$ [Eq. (\ref{eq:I_t_tot}) with Eq. (\ref{eq:I_t_m_cyl}) and (\ref{eq:I_t_p_cyl})  or 
(\ref{eq:I_t_m_sphere})  and (\ref{eq:I_t_p_sphere})] of the $e^+$ lifetime component $\tau_t$ of the interface-trapped state.
Both $\overline{\tau}$ and $I_t$ consist of volume-weighted parts associated with the precipitates (weighting factor: $(r_0/R)^i$, $i=2,3$)
and the matrix (weighting factor: $[1-(r_0/R)^i]$).\footnote{For instance for the cylindrical case the extraction of the weighting factor out of the bracket of the matrix term 
of $\overline{\tau}$ [Eq.~\ref{eq:tauq_cyl}]  yields: 
$\tau_f [1-(r_0/R)^2] [1+ 
2 \alpha r_0/(R^2-r_0^2)\, (\dots)]$. Analogous for the matrix part of $\overline{\tau}$ in 
the spherical case [Eq.~\ref{eq:tauq_sphere}]: 
$\tau_f [1-(r_0/R)^3] [1+ 
3 \alpha r_0^2/(R^3-r_0^3)\, (\dots)]$.
Likewise for the intensities of the matrix part [Eq.  (\ref{eq:I_t_m_cyl}) and  (\ref{eq:I_t_m_sphere})]
extraction of the weighting factor yields: \\
$[1-(r_0/R)^2] 2 \alpha r_0/(R^2-r_0^2)$ instead of $2 \alpha r_0/R^2$ [Eq.  (\ref{eq:I_t_m_cyl})]; 
$[1-(r_0/R)^3] 3 \alpha r_0^2/(R^3-r_0^3)$ instead of $3 \alpha r_0^2/R^3$ [Eq.  (\ref{eq:I_t_m_sphere})].}
%%%%%%%%%%%%%%%%%%%%%%%%%%%%%%%
Apart from the weighting factor [$(r_0/R)^3$], the precipitate part of $\overline{\tau}$ for the spherical case 
[Eq.  (\ref{eq:tauq_sphere})] is  identical to that deduced earlier for
$e^+$ trapping at grain boundaries of spherical crystallites \cite{Wuerschum96}. Likewise, 
the precipitate parts of  $I_t$ (without $(r_0/R)^i$, $i=2,3$)  
[Eq.  (\ref{eq:I_t_p_cyl}) and (\ref{eq:I_t_p_sphere})] 
are identical to those obtained for
$e^+$ trapping at grain boundaries of cylindrical \cite{dryzek1998} or spherical crystallites \cite{Wuerschum96, dryzek1998}. 

\begin{figure}
\includegraphics[width=7.5cm]{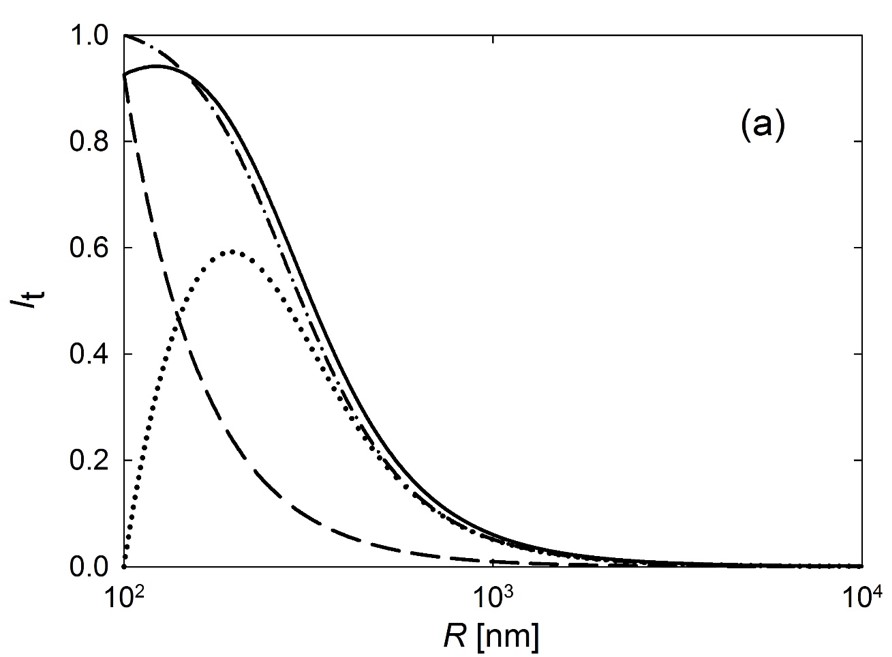}
\includegraphics[width=7.7cm]{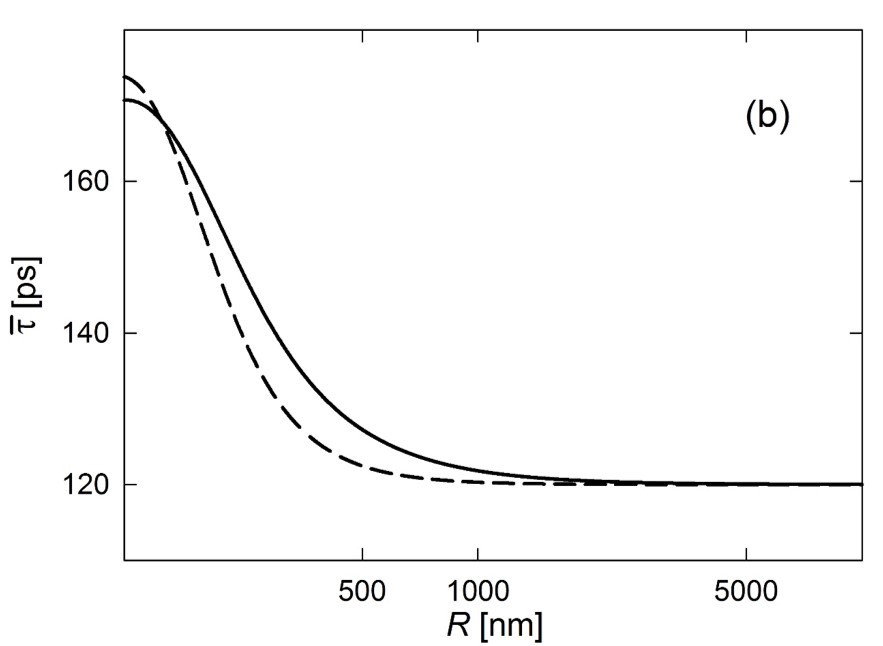}
\caption{(a) Relative intensity $I_t$ [Eq.~(\ref{eq:I_t_tot}), solid line] of interface-trapped component $\tau_t$
in dependence of diffusion radius $R$ for cylindrical-shaped precipitates.
The dashed and dotted line shows the part which arises from $e^+$ trapping from
the precipitate [$I_t^{precip}$, Eq.~(\ref{eq:I_t_p_cyl})] and from the matrix [$I_t^{matrix}$, Eq.~(\ref{eq:I_t_m_cyl})],
respectively. The dash-dotted line shows $I_t^{matrix}$ [Eq.~(\ref{eq:I_t_m_cyl})] without the weighting factor
$[1-(r_0/R)^2]$ [Eq. (\ref{eq:I_t_m_cyl_hollow})].
(b) Mean $e^+$ lifetime $\overline{\tau}$ in dependence of diffusion radius $R$
for precipitate$-$matrix composite
with cylindrical [Eq.~(\ref{eq:tauq_cyl}), solid line] or spherical symmetry [Eq.~(\ref{eq:tauq_sphere}), dashed line].
Parameters: $\tau_f= \tau_p = 120$~ps, $\tau_t=180$~ps,
$D=10^{-4}$~m$^2$s$^{-1}$, $\alpha=  \beta = 3 \times 10^3$~ms$^{-1}$,
$r_0 = 100$~nm. Note that $R$ is related to number density of precipitates [Eq. \ref{eq:N_p}].
}
\label{fig:2}
\end{figure}

Fig.~\ref{fig:2}.a shows the intensity  $I_t$ [Eq.~(\ref{eq:I_t_tot})] in dependence of diffusion radius $R$ for cylindrical-shaped precipitates of constant diameter $r_0$. With decreasing $R$, i.e., with increasing number of precipitates [Eq. \ref{eq:N_p}],
the characteristic sigmoidal-shaped increase of $I_t$ occurs. Also plotted in Fig.~\ref{fig:2}.a are the two parts 
of $I_t$ arising from the precipitate and matrix, i.e., $I_t^{precip}$ 
[Eq.~(\ref{eq:I_t_p_cyl})] and $I_t^{matrix}$ [Eq.~(\ref{eq:I_t_m_cyl})], respectively.
Due to the constant precipitate size, the increase of  $I_t^{precip}$ with decreasing $R$
exclusively reflects the variation of the weighting factor $(r_0/R)^2$.  Remarkably,  $I_t^{matrix}$ shows a maximum. 
The increase of $I_t^{matrix}$ with decreasing $R$ arises from the increasing trapping due to a decrease of the maximum 
$e^+$ diffusion length necessary for $e^+$ for reaching the interface. 
For small values of $R$ this  $I_t^{matrix}$ increase with decreasing $R$ due to the diffusion effect is counterbalanced by the 
effect of the weighting factor, i.e., the decreasing relative initial fraction of $e^+$ in the matrix compared to that in
precipitates. This is illustrated by the plot of $I_t^{matrix}$ without the weighting factor (see Fig.~\ref{fig:2}.a)
which shows the expected sigmoidal-shaped increase over the entire $R$ regime.\footnote{$I_t^{matrix}$
without the weighting factor corresponds to eq.~(\ref{eq:I_t_m_cyl_hollow}); see below.}

A comparison of cylindrical and spherical symmetry is shown in Fig.~\ref{fig:2}.b by the example of
the mean positron lifetime $\overline{\tau}$. Similar to $I_t$, $\overline{\tau}$ exhibits the 
characteristic sigmoidal-shaped increase with decreasing $R$, i.e., increasing precipitates concentration.
Over nearly the entire $R$-regime, the  $\overline{\tau}$-increase for cylindrical precipitates is higher than for spherical ones, which means that for a given $R$ the trapping rate at interfaces of cylindrical precipitates exceeds 
that of  spherical precipitates. This is simply because the trapping active area for 
cylinders is larger than for spheres for a given $R$.

Finally one should mention that the model can also be applied to extended open-volume defects like spherical voids
or hollow cylinders. In that case  $e^+$ trapping  occurs exclusively from the matrix outside 
the internal surface of the defect (rather than at the matrix$-$precipitate interface from both sides).
The solutions for the mean positron lifetime $\overline{\tau}$ and the trap intensity $I_t$
directly follows from the equations given above by omitting the part arising from the precipitate.
Assuming as initial condition that for $t=0$ no positrons are inside the trap,
the matrix weighting factor $[1-(r_0/R)^i] (i= 2,3)$ has to be replaced by 1.
For the  hollow cylinders the solutions read:\footnote{The solutions for voids were reported recently.\cite{Wuerschum18}
} [cf. Eq. (\ref{eq:tauq_cyl}), eq.  (\ref{eq:I_t_m_cyl}), and first footnote on previous page.] 
\begin{equation}
\label{eq:tauq_cyl_hollow}
\overline{\tau} = \tau_f
\Biggl\{1 + \frac{2 \alpha r_0}{R^2-r_0^2} (\tau_t - \tau_f)
\frac{\Lambda_1}{\Lambda_1 - (\tau_f/D)^{1/2} \alpha \Lambda_0} \Biggr\} \, 
\mbox{ with }  \Lambda_{0,1} = \Lambda_{0,1} (\gamma_0 r_0, \gamma_0 R) \,
,
\end{equation}
\begin{equation}
I_t = \frac{2 \alpha r_0}{R^2-r_0^2} \, 
\frac{\Lambda_1}{(\tau_f^{-1} - \tau_t^{-1}) \Lambda_1 - \gamma_t \alpha \Lambda_0} \, 
\mbox{ with }  \Lambda_{0,1} = \Lambda_{0,1} (\gamma_t r_0, \gamma_t R) \, .
\label{eq:I_t_m_cyl_hollow}
\end{equation}

In conclusion, the model presents the basis for studying all types
of cylindrical- and spherical-shaped extended defects irrespective of their size and  their number density.
Of particular relevance are matrix$-$precipitate composites for which such a model 
could be established in the present work.
An intriguing feature of the model are the closed-form solutions which can be conveniently applied for
the analysis of experimental data.

%%%%%%%%%%%%%%%%%%%%%%%%%%%%%%%%%%%%%%%%%%%%%%%%%%%%%%%%%%%%%%%%%%%%%%%%%%%%%%%
\section*{Appendix: Derivation of Laplace transform $\tilde{n}(p)$}
The appendix gives a brief summary on the solution of the diffusion and rate equations (\ref{eq:diffusion}, \ref{eq:rho_t})
for cylindrical precipitates resulting in the Laplace transform $\tilde{n}(p)$ [Eq. (\ref{Laplace_n_cyl})] (for spherical
precipitates see Ref. \cite{Wuerschum18}).
The time dependence of equations (\ref{eq:diffusion}) and (\ref{eq:rho_t})
is solved by means of Laplace transformation ($\rho_{m/p/t}(r,t) \rightarrow \tilde{\rho}_{m/p/t}(r,p)$)
yielding the modified Bessel differential equation for the $r$-dependence of $\tilde{\rho}_{m/p}(r,p)$.
Taking into account the initial conditions, the solutions of eq. (\ref{eq:diffusion}) and (\ref{eq:rho_t})  read 
\begin{equation}
\label{solution_rho}
\tilde{\rho}_{m/p} (r,p)=A \, I_{0}(\gamma r)+ B \, K_{0}(\gamma r) +
\frac{\rho_{m/p}(0)}{\tau_{f/p}^{-1} + p} \, , \quad
\tilde{\rho}^{(m)}_t=
\frac{\alpha \tilde{\rho}_m(r_{0},p)}{
\tau_{t}^{-1}+p} \, , \quad
\tilde{\rho}^{(p)}_t=
\frac{\beta \tilde{\rho}_p(r_{0},p)}{
\tau_{t}^{-1}+p} 
\end{equation}
with $\gamma^{2} = \gamma^{2}(p)= (\tau_{f/p}^{-1}+p)/D$ and the  modified Bessel functions
$I_{j}$, $K_{j}$ \cite{olver2010nist}.
The coefficients $A$ and $B$ as determined by the boundary conditions [Eq.~(\ref{eq:inner_bound}), 
eq.~(\ref{eq:outer_bound})]  read for the inner (p) and outer part (m):
\begin{equation}
A = \displaystyle{\frac{- \beta \rho_p(0)}{\tau_p^{-1}+p} \, \frac{1}{D \gamma I_1 (\gamma r_0) + \alpha I_1 (\gamma r_0)}} \, ,
B = 0
\mbox{ and } \quad 
A= \displaystyle{\frac{\alpha \rho_m(0)}{\tau_f^{-1}+p} \,\frac{K_1(\gamma R)}{D \gamma \Lambda_1 - \alpha \Lambda_0}} \, , \,
B= \displaystyle{\frac{\alpha \rho_m(0)}{\tau_f^{-1}+p} \,\frac{I_1(\gamma R)}{D \gamma \Lambda_1 - \alpha \Lambda_0}} 
\label{eq:AB}
\end{equation}
respectively, with $\Lambda_{0,1}$ according to eq.~(\ref{eq:L_12}).
From the densities $\tilde{\rho}_{m/p} (r,p)$,  
$\tilde{\rho}^{(m,p)}_t$ [Eq.~(\ref{solution_rho})] follows the Laplace transform $\tilde{n}(p)$ of  
the total probability $n(t)$ that a
$e^+$ implanted at $t=0$ has not yet been annihilated at time $t$. This is  obtained by integration over 
the cylindrical volume of the precipitate and over the hollow-cylindrical matrix. For the above mentioned initial 
condition ($\rho_m(0) = \rho_p(0)$)
$\tilde{n}(p)$ reads:
\begin{equation}
\tilde{n}(p)=
\frac{1}{\pi R^2 \rho_m(0)}
\left\{
 \int\limits^{r_0}_{0} 2\pi r \tilde{\rho}_p (r,p) {\rm d} r +
 \int\limits^{R}_{r_0} 2\pi r \tilde{\rho}_m(r,p) {\rm d} r
+ 2\pi r_0 \Bigl(\tilde{\rho}^{(m)}_t(p) + \tilde{\rho}^{(p)}_t(p)\Bigr)
\right\}  \,    
\end{equation}
which after  solving the integral results in eq.~(\ref{Laplace_n_cyl}).

%%%%%%%%%%%%%%%%%%%%%%%%%%%%%%%%%%%%%%%%%%%%%%%%%%%%%%%%%%%%%%%%%%%%%%%%%%%%%%%%%%%%%%%
%\section*{References}
%\nocite{*}
%\bibliographystyle{aipnum-cp}%

%\bibliography{wuerschum_trapping}
%Control: year  (1) truncated

\end{document}